\pdfoutput=1

\documentclass{PoS}

\usepackage{graphics}
\usepackage{graphicx}
\usepackage{color}
\usepackage{amsmath}
\usepackage{slashed}
\usepackage{epsfig}

\title{How to avoid unnatural hierarchical thermal leptogenesis}
\ShortTitle{How to avoid unnatural hierarchical thermal leptogenesis}

\author{\speaker{Jackson D. Clarke}%
         \thanks{Based on work completed in collaboration with Robert Foot and Raymond R. Volkas \cite{Clarke2015gwa,Clarke2015hta}.}\\
        ARC Centre of Excellence for Particle Physics at the Terascale, \\ 
School of Physics, University of Melbourne, 3010, Australia\\
        E-mail: \email{j.clarke5@pgrad.unimelb.edu.au}}
        
\abstract{A one-flavour naturalness argument suggests that
the Type I seesaw model cannot naturally explain
neutrino masses and the
baryon asymmetry of the Universe via
hierarchical thermal leptogenesis.
We prove that there is no way to avoid this
conclusion in a minimal three-flavour setup. 
We then comment on the simplest ways out.
In particular, we focus on a resolution utilising a second Higgs doublet.
Such models predict an automatically SM-like Higgs boson, 
(maximally) TeV-scale scalar states, 
and low- to intermediate-scale hierarchical leptogenesis 
with $10^3\text{ GeV}\lesssim M_{N_1}\lesssim 10^7\text{ GeV}$.}

\FullConference{18th International Conference From the Planck Scale to the Electroweak Scale \\
                 25-29 May 2015\\
                 Ioannina, Greece }

\begin{document}

\section{Introduction}

The standard model (SM) and the paradigm of electroweak symmetry breaking
realised by the Higgs potential
$V_{\text{SM}} = \mu^2\Phi^\dagger\Phi + \lambda \left(\Phi^\dagger\Phi\right)^2$,
with $\mu^2(m_Z) \approx -(88\text{ GeV})^2$, 
has been extremely successful in explaining low energy phenomena.
However it fails to explain neutrino masses
and the baryon asymmetry of the Universe (BAU).
A straightforward way to explain both is to add
three heavy right-handed neutrinos.
Gauge invariance then allows two additional renormalisable terms
in the Yukawa Lagrangian,
\begin{align}
 -\Delta\mathcal{L}_Y = (y_\nu)_{ij} \overline{l^i_L}\tilde\Phi\nu^j_R 
 +\frac12 M_i\overline{(\nu^i_R)^c}\nu^i_R +h.c., \label{EqSeesawYuk}
\end{align}
where $l_L=(\nu_L,e_L)^T$, $\tilde\Phi=i\tau_2\Phi^*$,
and $M_i$ are the right-handed neutrino masses.
This is the Type I seesaw model \cite{Minkowski1977sc,Mohapatra1979ia,Yanagida1979as,GellMann1980vs}.
After $\Phi$ gains a vacuum expectation value (vev) $\langle\Phi\rangle=v/\sqrt{2}\approx 174$~GeV,
and if $y_\nu v \ll M_i$, the neutrino mass matrix is given by the seesaw formula
\begin{align}
 m_\nu = \frac{v^2}{2} y_\nu \mathcal{D}^{-1}_M y_\nu^T, \label{EqSeesaw}
\end{align}
where $\mathcal{D}_M\equiv \text{diag}(M_1,M_2,M_3)$.
The BAU can be produced via hierarchical thermal leptogenesis \cite{Fukugita1986hr}:
the $CP$ violating out-of-equilibrium 
decays of the lightest right-handed neutrino $N_1$
create a lepton asymmetry which is transferred to the baryon
sector by electroweak sphalerons.
The Davidson--Ibarra bound \cite{Davidson2002qv,Giudice2003jh} (ensuring enough $CP$ violation)
for successful hierarchical ($M_{N_1}\ll M_{N_2}\ll M_{N_3}$) thermal leptogenesis is
\begin{align}
 M_{N_1} \gtrsim 5\times 10^8\text{ GeV} \left(\frac{v}{246\text{ GeV}}\right)^2, \label{EqDavIbarra}
\end{align}
where $v$ is the vev that enters the seesaw Eq.~\ref{EqSeesaw}.
This appears to be in conflict with the naturalness argument
for right-handed neutrinos made by Vissani \cite{Vissani1997ys}.
In a \textit{one-flavour} model, Vissani found (where $\mu_R$ is the renormalisation scale)
\begin{align}
 \left|\frac{d\mu^2}{d\ln\mu_R}\right| = \left|-\frac{1}{4\pi^2} y_\nu M_N^2 y_\nu^*\right| 
  < 1\text{ TeV}^2 && \Rightarrow && M_N \lesssim 3\times 10^7\text{ GeV}\left(\frac{v}{246\text{ GeV}}\right)^\frac23  \label{EqNatural}
\end{align}
for a neutrino mass of $m_\nu = \frac{v^2}{2}\frac{y_\nu^2}{M_N} \approx 0.05$~eV.

This proceedings paper addresses the following questions:
can three-flavour effects ameliorate this conflict?
and; if not, what can?
In Sec.~\ref{Sec3Flav} the first question is answered in the negative.
We outline our three-flavour treatment \cite{Clarke2015gwa} 
which generalises the Vissani result 
to obtain three naturalness bounds:
\begin{align}
 M_{N_1} \lesssim 4\times 10^7 \text{ GeV},&& 
 M_{N_2} \lesssim 7\times 10^7 \text{ GeV},&& 
 M_{N_3} \lesssim 3\times 10^7 \text{ GeV}\left(\frac{0.05\text{ eV}}{m_{min}}\right)^\frac13, \label{eqboundN}
\end{align}
where $m_{min}$ is the lightest neutrino mass.
These results confirm that natural
$N_1$-, $N_2$-, or $N_3$-dominated hierarchical thermal leptogenesis
is not possible in a minimal three-flavour Type I seesaw.
In Sec.~\ref{Secnu2HDM} we suggest some simple variations/extensions 
which reopen the possibility of a natural BAU.
We focus on a two-Higgs-doublet solution recently
proposed in our Ref.~\cite{Clarke2015hta},
motivated by the following observation: 
if $v\lesssim 30$~GeV in Eq.~\ref{EqSeesaw},
then Eqs.~\ref{EqDavIbarra} and \ref{EqNatural} become compatible.
We find viable natural models which
predict a SM-like Higgs boson, 
(maximally) TeV-scale scalar states, and
low- to intermediate-scale hierarchical leptogenesis 
with $10^3\text{ GeV}\lesssim M_{N_1}\lesssim 10^8\text{ GeV}$.

\section{Electroweak naturalness in three-flavour Type I seesaw \label{Sec3Flav}}

\subsection{Measurable naturalness}

After renormalisation, the physical effects of any 
heavy degree of freedom are embodied in the renormalisation group equations (RGEs).
The RGEs are therefore a sensible way to quantify a 
physical and \textit{measurable} (at least in principal) 
electroweak naturalness problem.
Roughly, a problem arises whenever $d\mu^2/d\ln\mu_R \gtrsim \mu^2$;
in such a case, $\mu^2(\mu_R)$ will evolve to large values,
which one can interpret as a fine-tuning of $\mu^2$ at a high scale.
Intuitive naturalness criteria are then:
bound the RGE itself, or;
quantify and bound the fine-tuning 
in the mass parameter evolved to some high scale $\Lambda_h$. That is:
\begin{align}
 \left|\frac{1}{\mu^2(m_Z)}\frac{d\mu^2}{d\ln\mu_R}\right| < \mathcal{O}(1) ,
 && \text{or;} &&
 \Delta\left(\Lambda_h\right) = \left| \frac{\mu^2(\Lambda_h)}{\mu^2(m_Z)}
  \frac{\partial \mu^2(m_Z)}{\partial \mu^2(\Lambda_h)} \right| 
  < \mathcal{O}(1), \label{EqDelta}
\end{align}
where $\Delta$ is a Barbieri-Giudice style fine-tuning measure \cite{Ellis1986yg,Barbieri1987fn}.
Such criteria should not be taken too seriously
(and nature may just be fine-tuned after all),
but they can certainly serve as guiding principles which 
capture our subjective sense of
physical naturalness (of mass parameters),
and they are calculable in any perturbative model.

\subsection{Three-flavour seesaw}

\begin{figure}[t]
 \centering
 \includegraphics[width=0.48\columnwidth]{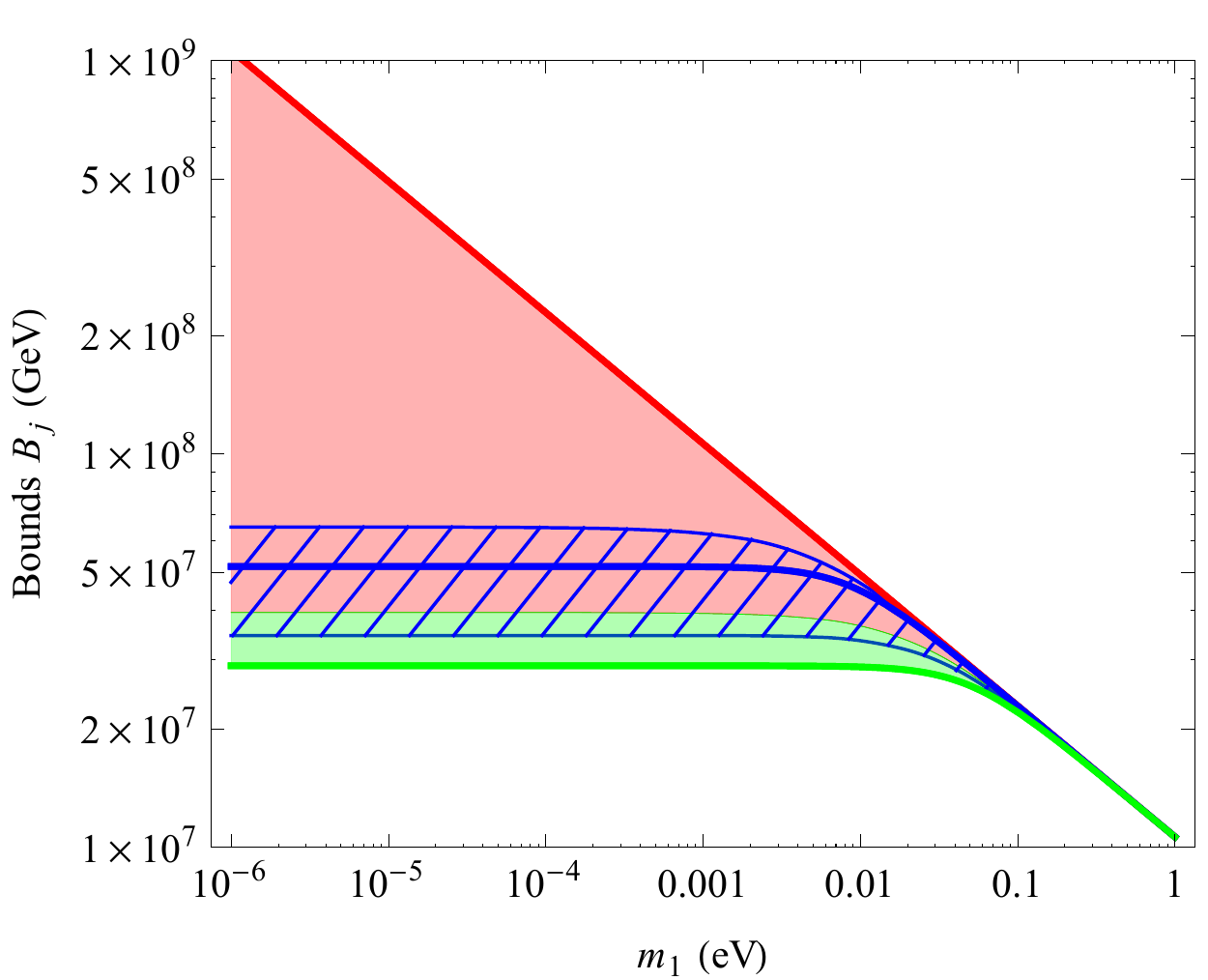}
 \includegraphics[width=0.48\columnwidth]{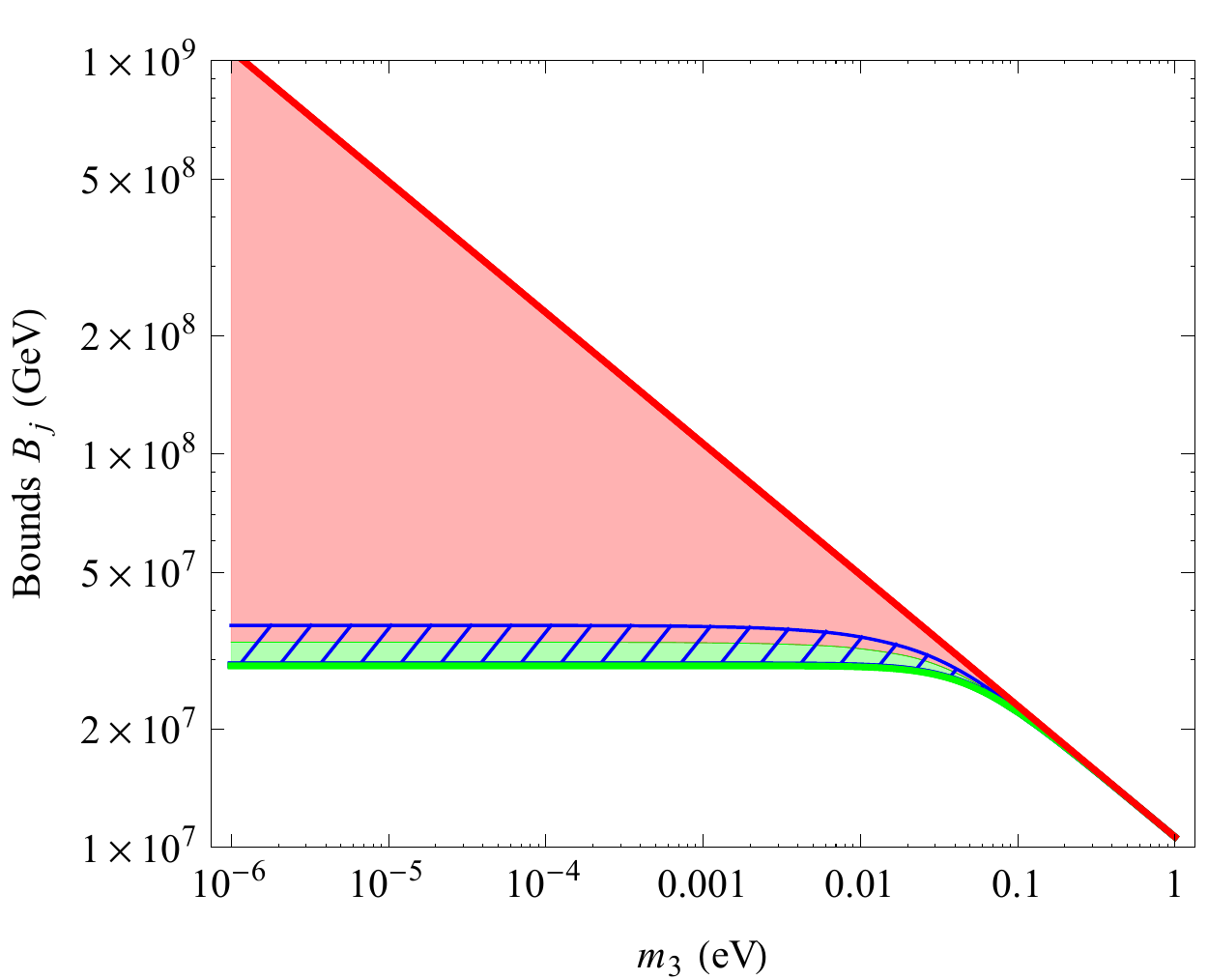}
 \caption{Left: As a function of the lightest neutrino mass in normal ordering,
 shown as red/blue-hatched/green (darker/hatched/lighter)
 are the attainable values for $B_3\ge B_2\ge B_1$ 
 naturalness bounds on the $M_{N_i}$, 
 by requiring $d\mu^2/d\ln\mu_R<1\text{ TeV}^2$.
 The regions assume $R$ is real.
 Thick solid lines show the case $R=\mathbb{I}$
 (the complex $R$ case is similar).
 Right: As in Left but for inverted ordering. 
 Note that the thick blue line is obscured by the thick green line.
 (Figure reproduced from Ref.~\cite{Clarke2015gwa}).}
 \label{fig3Flavbounds}
\end{figure}

Let us now examine right-handed neutrino corrections 
to the electroweak $\mu^2$ parameter 
in the three-flavour Type I seesaw model.
We will invoke the Casas-Ibarra parameterisation 
$U y_\nu = \frac{\sqrt{2}}{v} \mathcal{D}_m^\frac12 R \mathcal{D}_M^\frac12$,
where $R$ is an arbitrary unitary matrix, and
$\mathcal{D}_m\equiv \text{diag}(m_1,m_2,m_3)=Um_\nu U^T$
is the diagonalised neutrino mass matrix. 
The RGE for $\mu^2$ is
\begin{align}
  \frac{d\mu^2}{d\ln\mu_R} = \frac{1}{(4\pi)^2} 
  \left[ -4 \text{Tr}[y_\nu\mathcal{D}_M^2 y_\nu^\dagger] + \mathcal{O}(\mu^2) \right]
  = \frac{1}{(4\pi)^2} 
  \left[ -4 \frac{2}{v^2} \text{Tr}[\mathcal{D}_mR \mathcal{D}_M^3 R^\dagger] + \mathcal{O}(\mu^2) \right]. \label{Eqmu2sqRGE}
\end{align}
Bounding directly each right-handed neutrino contribution by $1\text{ TeV}^2$
(akin to Vissani) results in three bounds:
\begin{align}
 M_j \lesssim 3\times 10^7 \text{ GeV}
   \left(\frac{v}{246\text{ GeV}}\right)^\frac23 
   \left( \frac{0.05\text{ eV}}{\sum_i m_i |R_{ij}|^2} \right)^\frac13 , \label{eqdmult1TeV}
\end{align}
where $R_{ij}$ are the entries of $R$.
We can always order the bounds by their size;
we will call them $B_j$ and take $B_1\le B_2\le B_3$.
The question we are interested in is: 
what values of $B_j$ are attainable from Eq.~\ref{eqdmult1TeV}?
To answer this question we need only extremise the $B_j$ over $R$.
After a suitable parameterisation of $R$ and a numerical study
we present our result for real $R$ in Fig.~\ref{fig3Flavbounds},
as a function of the lightest neutrino mass in normal ordering 
($m_1<m_2<m_3$) and inverted ordering ($m_3<m_1<m_2$) scenarios.
The result for complex $R$ with $|R_{ij}|<1$ (to avoid a fine-tuning) is similar.
One can now plainly observe the generic naturalness bounds already 
written in Eq.~\ref{eqboundN}.

What are the implications for leptogenesis?
The Davidson--Ibarra bound (Eq.~\ref{EqDavIbarra}) for 
$N_1$-dominated thermal hierarchical leptogenesis remains inconsistent with naturalness.
An $N_2$-dominated scenario is also inconsistent \cite{DiBari2005st,Vives2005ra,Engelhard2006yg}.
Lastly, it turns out that the same decoupling limit which allows 
$N_3$ to become naturally heavy also sends the $CP$ asymmetry in its decays to zero,
excluding the possibility of a natural $N_3$-dominated scenario.
Thus our results confirm that no minimal three-flavour Type I seesaw model 
can explain the neutrino masses
and baryogenesis via hierarchical thermal leptogenesis 
while remaining completely natural.

\section{Natural leptogenesis and neutrino masses with two Higgs doublets: the $\nu$2HDM \label{Secnu2HDM}}

\subsection{How to avoid unnatural hierarchical thermal leptogenesis}

An obvious question is: in what minimal ways can we adapt the
Type I seesaw to realise a natural BAU?
There are a number of conspicuous possibilities:
(1) lowering the Davidson--Ibarra bound, 
by considering dominant initial $N_1$ abundancy \cite{Giudice2003jh}\footnote{The bound becomes
$M_{N_1}\gtrsim 2\times 10^7$~GeV, marginally consistent with 
the naturalness bound in Eq.~\ref{eqboundN}.},
resonant leptogenesis \cite{Pilaftsis2003gt},
a different baryogenesis mechanism entirely 
(such as neutrino oscillations \cite{Akhmedov1998qx}),
or by introducing new fields which allow 
increased $CP$ violation in $N_1$ decays;
(2) raising the naturalness bound by
partially cancelling right-handed neutrino corrections \cite{Bazzocchi2012de,Fabbrichesi2015zna},
or removing it entirely by restoring low-scale supersymmetry;
(3) lowering the (possibly effective) vev entering the seesaw Eq.~\ref{EqSeesaw}
so that the bounds of Eqs.~\ref{EqDavIbarra} and \ref{EqNatural} become consistent ($v\lesssim 30$~GeV).
Recently in Ref.~\cite{Clarke2015hta} we
implemented the latter possibility within a 
two-Higgs-doublet model
with right-handed neutrinos ($\nu$2HDM).
The remainder of this Section is dedicated to describing such models.

\subsection{The $\nu$2HDM}

There are two doublets $\Phi_{1,2}$ with hypercharge $+1$,
and each gains a non-zero vev $\langle \Phi_i\rangle=(0,v_i/\sqrt{2})^T$
with $\sqrt{v_1^2+v_2^2}=v\approx 246$~GeV and $\tan\beta\equiv v_1/v_2$.
As already motivated, we would like the vev contributing to the seesaw to be small.
We therefore consider $v_2\ll v_1$ ($\tan\beta \gg 1$) and the following Yukawa Lagrangian:
\begin{align}
 -\mathcal{L}_Y =&\, + y_u\overline{q_L}\tilde\Phi_1 u_R + y_d\overline{q_L}\Phi_I d_R
    + y_e \overline{l_L}\Phi_J e_R + y_\nu \overline{l_L}\tilde\Phi_2 \nu_R
    + \frac12 M_N \overline{(\nu_R)^c}\nu_R + h.c. , \label{EqYuk}
\end{align}
where $I,J$ define the $\nu$2HDM Type, and family indices are implied.
The model Types are defined in Table~\ref{TabTypes}.
Note that, for Type II, LS, and Flipped arrangements, $y_{b,\tau}$
cause early Landau poles when $v_2\lesssim 4$~GeV ($\tan\beta \gtrsim 70$).

In order to construct a model with naturally small $v_2$ and potentially TeV-scale scalars,
we softly break a symmetry which would otherwise imply $v_2=0$.
For example, take the softly broken $U(1)$ symmetric potential
\begin{align}
 V_{\text{2HDM}} = &\ 
m^2_{11}\, \Phi_1^\dagger \Phi_1
+ m^2_{22}\, \Phi_2^\dagger \Phi_2 
-m^2_{12}\, \left(\Phi_1^\dagger \Phi_2 + \Phi_2^\dagger \Phi_1\right)
\nonumber\\
&+ \frac{\lambda_1}{2} \left( \Phi_1^\dagger \Phi_1 \right)^2
+ \frac{\lambda_2}{2} \left( \Phi_2^\dagger \Phi_2 \right)^2
+ \lambda_3\, \left(\Phi_1^\dagger \Phi_1\right) \left(\Phi_2^\dagger \Phi_2\right)
+ \lambda_4\, \left(\Phi_1^\dagger \Phi_2\right) \left(\Phi_2^\dagger \Phi_1\right), \label{EqV2HDM}
\end{align}
with $m_{11}^2<0$, $m_{22}^2 >0$, and $m_{12}^2/m_{22}^2\ll 1$.
In this case
\begin{align}
 v_2 \approx \frac{1}{1+\frac{v_1^2}{2m_{22}^2}(\lambda_3+\lambda_4)} \frac{m_{12}^2}{m_{22}^2}v_1 , &&
 v_1 = \sqrt{\frac{2}{\lambda_1} \left[ 
   \frac{1}{\tan^2\beta}\left( m_{22}^2 +\frac12 \lambda_2 v_2^2\right) -m_{11}^2 
   \right]}.
\end{align}
In the limit $m_{22}^2\gg v_1^2(\lambda_{3}+\lambda_4)$, $\lambda_2 v_2^2$,
we have $\tan\beta\approx m_{22}^2/m_{12}^2$ and
$v_1 \approx \sqrt{\frac{2}{\lambda_1}\left( -m_{11}^2 + m_{12}^2 \right)}$.
This implies a relevant consistency condition,
$2m_{12}^2\lesssim \lambda_1 v_1^2$,
to ensure $m_{11}^2<0$ and avoid a fine-tuning for $v$.
Typically we have $m_{12}^2\ll |m_{11}^2|$ so that $m_{11}^2$ sets the
mass of the Higgs (as does $\mu^2$ in the SM).

\begin{table}[t]
 \centering
 \begin{tabular}{c|c|c|c|c}
  Model 	& $u_R^i$ 	& $d_R^i$ 	& $e_R^i$ 	& $\nu_R^i$ \\
  \hline
  Type I	& $\Phi_1$	& $\Phi_1$	& $\Phi_1$	& $\Phi_2$ \\
  Type II	& $\Phi_1$	& $\Phi_2$	& $\Phi_2$	& $\Phi_2$ \\
  Lepton-specific (LS)	& $\Phi_1$	& $\Phi_1$	& $\Phi_2$	& $\Phi_2$ \\
  Flipped	& $\Phi_1$	& $\Phi_2$	& $\Phi_1$	& $\Phi_2$ \\
 \end{tabular}
 \caption{The $\nu$2HDM Types.}
 \label{TabTypes}
\end{table}

The lightest $CP$ even boson $h$ obtains a mass $m_h^2\approx \lambda_1 v_1^2$.
Because of the approximate $U(1)$ symmetry,
a notable side effect is an automatically SM-like $h$,
i.e., there is no fine-tuning in the mixings to reproduce observations.\footnote{The 
relevant quantity is $\cos(\alpha-\beta)\sim \frac{1}{\tan\beta}\frac{v_1^2}{m_{22}^2}\ll 1$, see Ref.~\cite{Clarke2015hta}.}
The three extra scalar states ($H,A,H^\pm$) have masses $\approx m_{22}$.
Important constraints on $m_{22}$ in a given $\nu$2HDM Type 
are largely identical to those for a 2HDM of the same Type.
These are: 
the consistency condition already mentioned;
$m_{H^\pm}\gtrsim 80$~GeV from direct searches at LEP \cite{Searches2001ac};
$m_{22}\gtrsim 480$~GeV (for Type II and Flipped) 
from radiative $B\to X_s \gamma$ decays \cite{Amhis2014hma,Misiak2015xwa};
from $H/A\to \tau\tau$ LHC searches \cite{Aad2014vgg,Khachatryan2014wca}
a bound (for Type II) rising approximately linearly from 
$m_{22}\gtrsim 300$~GeV at $\tan\beta=10$ to $m_{22}\gtrsim 1000$~GeV at $\tan\beta=60$, and;
early Landau poles (for Type II, Flipped, and LS models) when $\tan\beta \gtrsim 70$ \cite{Bijnens2011gd}.

\subsection{Neutrino masses and leptogenesis}

\begin{figure}[t]
 \centering
 \includegraphics[width=0.48\columnwidth]{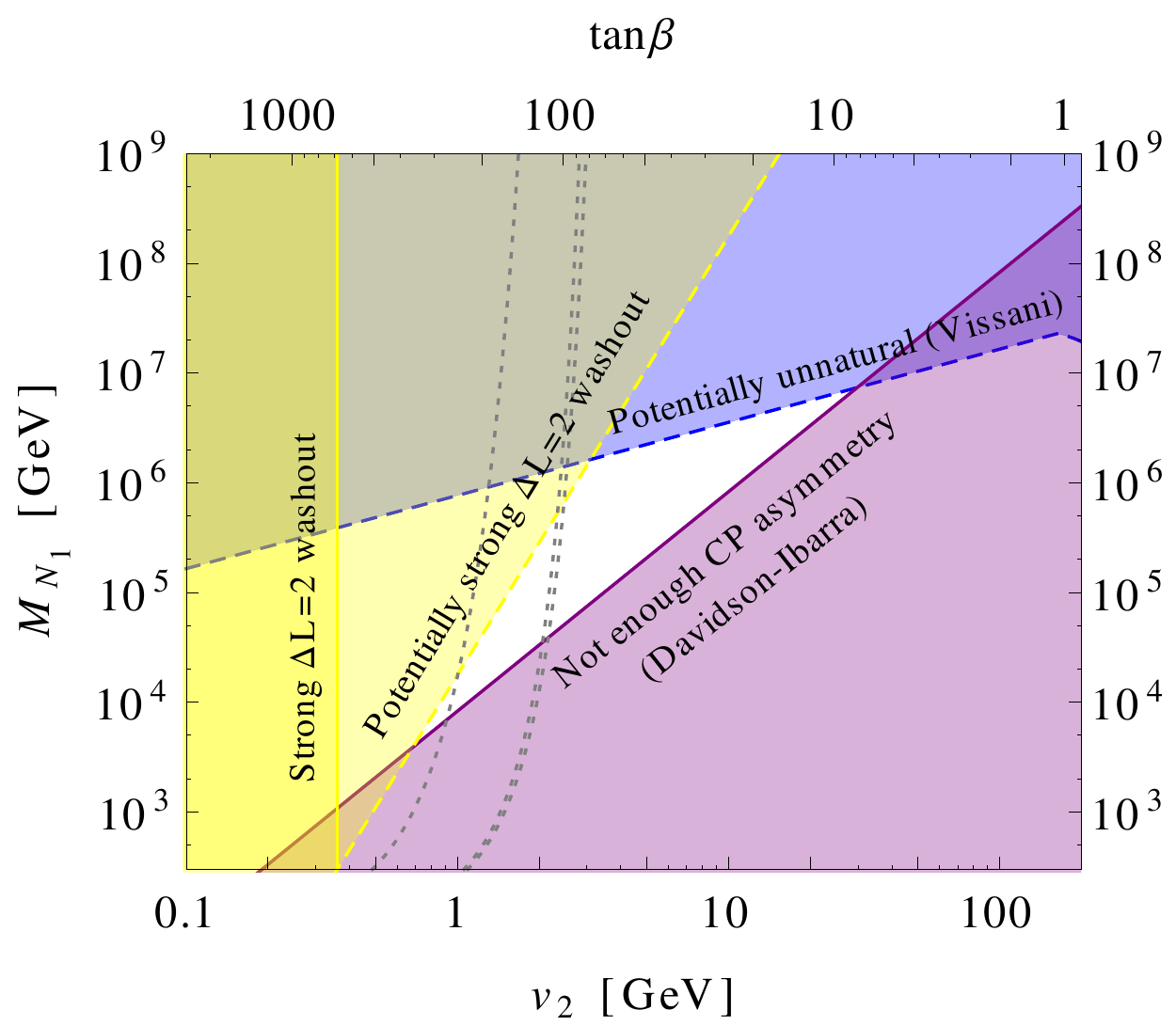}
 \includegraphics[width=0.48\columnwidth]{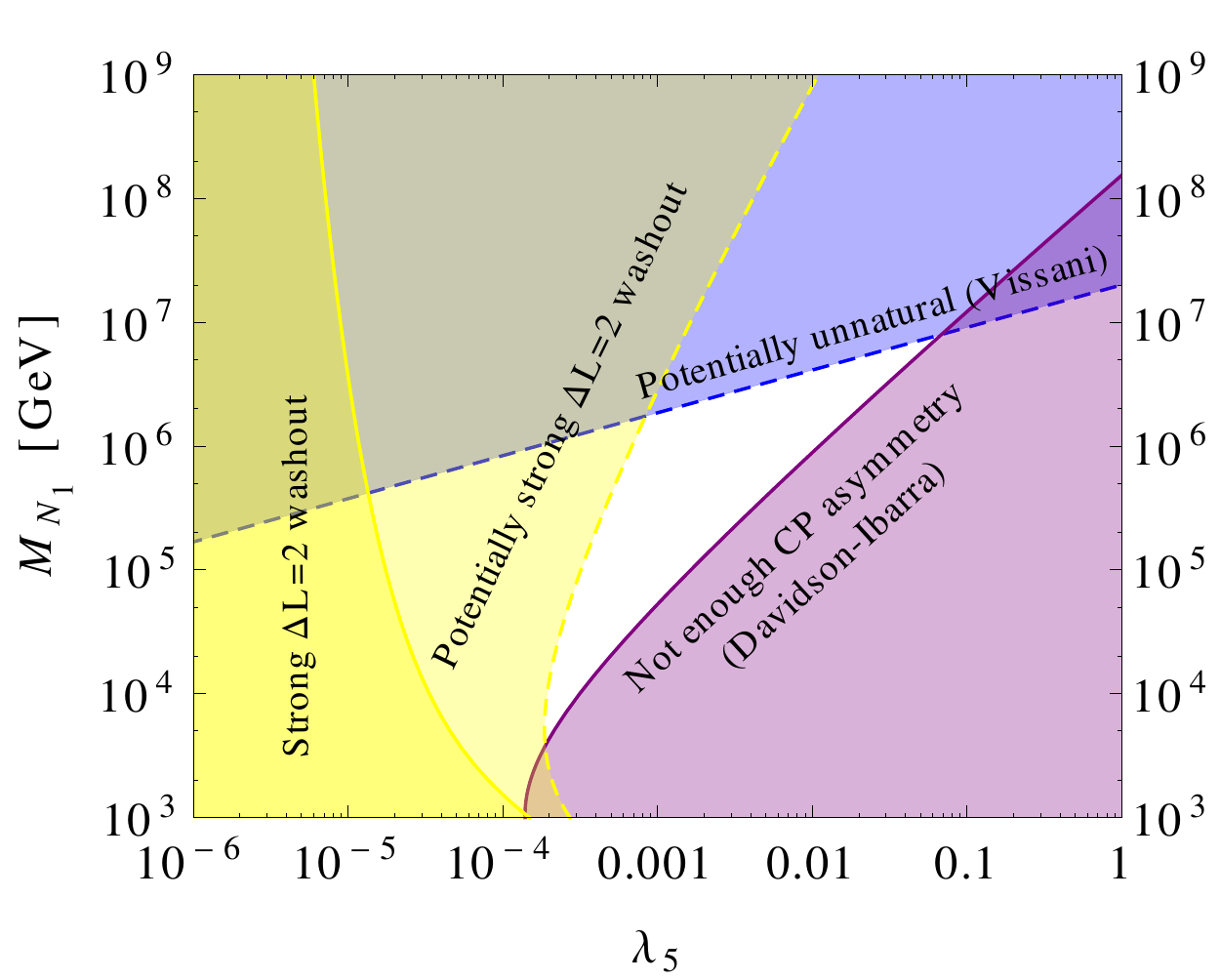}
 \caption{Left: Bounds on the $\nu$2HDM as a function of $v_2$.
 Shown (as labelled) are the Davidson-Ibarra bound,
 the Vissani naturalness bound, 
 and the areas of parameter space with strong $\Delta L=2$ scattering washout.
 The grey dotted lines indicate the $v_2$ below which 
 the Yukawas hit a Landau pole before $M_{N_1}$
 in the Type II, Flipped, and LS $\nu$2HDMs right-to-left.
 Right: As in Left but for the Ma model. 
 Each bound is evaluated at $m_{22}=500$~GeV.
 (Figure adapted from Ref.~\cite{Clarke2015hta}).}
 \label{Fignu2HDM}
\end{figure}

The neutrino mass matrix is given by
\begin{align}
 m_\nu = \frac{v_2^2}{2} y_\nu \mathcal{D}_M^{-1} y_\nu^T 
   \approx \frac{1}{\tan^2\beta} \left( \frac{v^2}{2} y_\nu \mathcal{D}_M^{-1} y_\nu^T \right), \label{Eqnumass2HDM}
\end{align}
suppressed with respect to the standard seesaw Eq.~\ref{EqSeesaw}.
Clearly a smaller $v_2$ forces a larger $y_\nu$ in order to realise the observed neutrino masses,
and it is the size of the $y_\nu$ entries which control the amount of 
$CP$ violation present in $N_1$ decays.
As such, and if leptogenesis proceeds in a sufficiently similar way
(we will soon discuss that it does),
the Davidson--Ibarra bound Eq.~\ref{EqDavIbarra} 
is suppressed by $\approx 1/\tan^2\beta$
and the scale of successful leptogenesis can be lowered.
This is illustrated in Fig.~\ref{Fignu2HDM}.

The observed BAU is produced analogously to standard
hierarchical thermal leptogenesis
(see e.g. Refs.~\cite{Buchmuller2004nz,Davidson2008bu} for reviews),
via the out-of-equilibrium, $CP$ violating decays 
of the lightest right-handed neutrino, 
but now into the second Higgs doublet: $N_1 \to l \Phi_2$.
When only decays and inverse decays are considered,
and in the one-flavour approximation,
the decay parameter $K$ characterises the asymmetry:
\begin{align}
 K = \frac{\Gamma_D}{H|_{T=M_1}} = \frac{\tilde m_1}{m_*},
\end{align}
where $\Gamma_D$ is the $N_1$ decay rate, 
$H$ is the expansion rate of the Universe,
$\tilde m_1$ is the effective neutrino mass,
and $m_*$ is the equilibrium neutrino mass,
\begin{align}
 \Gamma_D = \frac{1}{8\pi} (y_\nu^\dagger y_\nu)_{11} M_1, &&
 H  \approx \frac{ 17 T^2 }{M_{Pl}}, &&
 \tilde m_1 \equiv \frac{ (y_\nu^\dagger y_\nu)_{11} v_2^2}{2 M_1}, &&
 m_*  \approx \frac{ 1.1\times 10^{-3} \text{ eV} }{\tan^2\beta} . \label{Eqmtildemstar}
\end{align}
With these definitions we have the familiar weak and strong
washout regimes when $K\ll 1$ and $K\gg 1$, respectively.
Note that $m_*$ is smaller than its usual value in standard leptogenesis.

The $2\leftrightarrow 2$ scatterings with $\Delta L=1$
are important for washout and early $N_1$ production in the non-thermal weak washout regime.
Electroweak scatterings are identical to those in the standard scenario,
however those involving top quarks 
($Nl \leftrightarrow tq, Nt \leftrightarrow lq, Nq \leftrightarrow lt$)
are absent by construction.
Instead, at sufficiently large $\tan\beta$, 
top scatterings are replaced by the analogous bottom quark and tau lepton scatterings
(depending on the $\nu$2HDM type).\footnote{A large 
$y_\tau$ also introduces new scattering diagrams:
$N\Phi_2 \leftrightarrow \tau\Phi_2, \tau N \leftrightarrow \Phi_2\bar{\Phi}_2$.}
All of these scatterings are proportional to
$(y_\nu^\dagger y_\nu)_{11}$,
as are the decays and inverse decays,
so that they can only result in 
a minor departure from the standard scenario.

The $2\leftrightarrow 2$ scatterings with $\Delta L = 2$ 
($\Phi_2 l \leftrightarrow \bar{\Phi}_2\bar{l}, \Phi_2\Phi_2 \leftrightarrow ll$)
\textit{can} however have a much larger impact.
The rate of these scatterings is proportional to 
Tr$[(y_\nu y_\nu^T) (y_\nu y_\nu^T)^\dagger]$;
comparing this to the rate of decays, inverse decays, and $\Delta L = 1$ scatterings, we have
\begin{align}
 \frac{\text{Tr}[(y_\nu y_\nu^T) (y_\nu y_\nu^T)^\dagger]}{(y_\nu^\dagger y_\nu)_{11}}
 \propto \frac{M_{N_1}^2 \overline{m}^2 }{v_2^4} \frac{v^2}{2 M_{N_1} K \times 10^{-3}\text{ eV}} ,
\end{align}
where $\overline{m}^2=\sum{m_i^2}\gtrsim (0.05$~eV$)^2$.
Clearly this ratio increases as $v_2$ decreases ($\tan\beta$ increases).
For $T\lesssim M_{N_1}/3$ the $\Delta L = 2$ scattering rate 
is approximated (in the one-flavour approximation) by \cite{Buchmuller2004nz}
\begin{align}
 \frac{\Gamma_{\Delta L=2}}{H} \approx \frac{T}{2.2\times 10^{13}\text{ GeV}}
  \left(\frac{\overline{m}}{0.05\text{ eV}}\right)^2 \left(\frac{v}{v_2}\right)^4 , \label{EqDL2Scat}
\end{align}
In Fig.~\ref{Fignu2HDM} we show two regions of interest for these scatterings: 
when they are in equilibrium at $T\lesssim M_{N_1}/3$ and $T\sim 100$~GeV.
In these regions strong $\Delta L = 2$ washout 
can potentially destroy any asymmetry created.\footnote{Note that these
regions are not applicable in the non-perturbative regime
indicated by the grey dotted lines in Fig.~\ref{Fignu2HDM}.}

Lastly we note that, since natural leptogenesis will generically
be occurring at $T< 10^9$~GeV,
flavour effects cannot be ignored 
(see e.g. Refs.~\cite{Abada2006fw,Nardi2006fx,Abada2006ea,Blanchet2006be}).
It is known, for example, that flavour alignments can protect
the asymmetry from washout \cite{Vives2005ra}.
It is therefore plausible that successful leptogenesis is still possible
in the strong $\Delta L = 2$ washout region.
These effects deserve further study.
Still, the overall picture should not dramatically change,
and the (rescaled) Davidson--Ibarra bound is expected to hold 
(as it does the standard case with flavour effects \cite{Blanchet2006be,JosseMichaux2007zj}).

\subsection{Naturalness}

There are three explicit scales in the $\nu$2HDM: 
$m_{11}^2$, $m_{22}^2$, and the $M_j$.
A natural scenario is achieved if
(1) $m_{11}^2$ is protected from $m_{22}^2$ corrections, and
(2) $m_{22}^2$ is protected from $M_j$ corrections.
We consider each in turn.

In the typical situation where $m_{12}^2\ll |m_{11}^2|$,
$m_{11}^2$ sets the Higgs mass ($m_h^2\approx -2 m_{11}^2$)
so that $m_{11}^2\approx -(88\text{ GeV})^2$.
Naturalness considerations will imply,
because both $\Phi_i$ have gauge charges,
the $m_{22}^2$ scale cannot be very much separated from $m_{11}^2$.
At one-loop, the $m_{11}^2$ RGE is
\begin{align}
 \frac{dm_{11}^2}{d\ln\mu_R} = \frac{1}{(4\pi)^2} \left[ (4\lambda_3+2\lambda_4)m_{22}^2 + \mathcal{O}(m_{11}^2) \right]. \label{Eqm11sqRGE}
\end{align}
It appears that the limit $\lambda_{3,4}\to 0$ protects $m_{11}^2$ from $m_{22}^2$,
however these couplings are reintroduced by gauge loops:
\begin{align}
 \frac{d\lambda_3}{d\ln\mu_R} &= \frac{1}{(4\pi)^2} \left[ \frac34\left(g_Y^4-2g_Y^2g_2^2+3g_2^4\right) + ... \right],  &&
 \frac{d\lambda_4}{d\ln\mu_R} &= \frac{1}{(4\pi)^2} \left[ 3g_Y^2g_2^2 + ... \right]. \label{Eqlam34RGEs}
\end{align}
This is another way of saying there exists an irremovable pure-gauge two-loop
correction to $m_{11}^2$ that is proportional to $m_{22}^2$ 
(see Ref.~\cite{Chowdhury2015yja} for the two-loop result).
Bounding directly just the two-loop pure-gauge contribution by $1\text{ TeV}^2$ 
results in a conservative naturalness bound of $m_{22}\lesssim 10^5$~GeV.
It is also illuminating to consider the condition $\Delta(M_{Pl})<10$;
in Ref.~\cite{Clarke2015hta} we showed that this implies
a stringent naturalness bound $m_{22}\lesssim \text{few}\times10^3$~GeV.
These bounds are not to be taken too seriously,
they are merely sufficient to demonstrate that a TeV-scale $m_{22}$
is not only experimentally allowed in all $\nu$2HDM Types,
but can also remain natural.

Bounding the $m_{22}^2$ RGE directly by $dm_{22}^2/d\ln\mu_R < 1\text{ TeV}^2$
results in naturalness bounds on the $M_j$ given by
Eq.~\ref{eqdmult1TeV} with the replacement $v\to v_2$.\footnote{Actually, 
the Vissani bound may be relaxed
since $m_{22}$ can be naturally TeV-scale.
For clarity we will not consider this here;
see our Ref.~\cite{Clarke2015hta} for more details.}
This bound is depicted in Fig.~\ref{Fignu2HDM}.
A similar bound is obtained for $m_{22}\sim 1$~TeV and
a fine-tuning criterion $\Delta(M_{Pl})<10$.
We can now read off the region of parameter space of interest 
for natural leptogenesis:
we find, depending on the $\nu$2HDM Type,
fully perturbative solutions with $0.3\lesssim v_2/\text{GeV} \lesssim 30$
and $10^3\lesssim M_{N_1}/\text{GeV}\lesssim 10^7$.

\subsection{The Ma model}

Lastly let us comment that our discussion extends analogously to the 
Ma model of radiative neutrino mass \cite{Ma2006km}.
In this model the 2HDM potential is given by Eq.~\ref{EqV2HDM} with $m_{12}^2=0$
and an additional explicit $U(1)$ breaking term 
$\frac{\lambda_5}{2} [
( \Phi_1^\dagger\Phi_2 )^2
+ ( \Phi_2^\dagger\Phi_1 )^2 ]$,
retaining a $Z_2$ symmetry which remains unbroken ($v_2=0$).
For $M_N^2\gg m_{22}^2,v^2$,
the radiatively generated neutrino mass matrix is
\begin{align}
 (m_\nu)_{ij} \approx \frac{v^2}{2} \frac{(y_\nu)_{ik} (y_\nu^T)_{kj}}{M_k} \frac{\lambda_5}{8\pi^2} 
 \left( \ln\left[ \frac{2M_k^2}{(m_H^2+m_A^2)}\right] -1 \right). \label{EqnumassMa}
\end{align}
Arguments analogous to those already presented,\footnote{See
Ref.~\cite{Clarke2015hta} and Appendices therein
for some minor caveats}
but with $v_2^2\to v^2 \frac{\lambda_5}{8\pi^2} \left( \ln\left[ 2 M_{N_1}^2 / (m_H^2+m_A^2) \right] -1 \right)$
(c.f. Eqs.~\ref{EqnumassMa} and \ref{Eqnumass2HDM})
lead to a rescaling of the Davidson--Ibarra bound, 
the Vissani bound, and the strong $\Delta L=2$ scattering regions.
These bounds are shown in Fig.~\ref{Fignu2HDM} for an example mass $m_{22}=500$~GeV
(but they are only mildly dependent on $m_{22}$).
The region $10^{-5}\lesssim \lambda_5\lesssim 10^{-1}$
with $10^3\lesssim M_{N_1}/\text{GeV}\lesssim 10^7$
can naturally realise neutrino masses and hierarchical leptogenesis.
As well, $H$ or $A$ is a viable dark matter candidate.

\section{Conclusion}

The three-flavour Type I seesaw model is a simple way to explain neutrino masses 
and the BAU via hierarchical thermal leptogenesis.
However, as we proved in Sec.~\ref{Sec3Flav}, it cannot do so 
without introducing a naturalness problem \cite{Clarke2015gwa}.
In Sec.~\ref{Secnu2HDM} we listed some minimal ways 
to adapt the model to avoid this inconsistency:
dominiant initial $N_1$ abundancy; 
resonant leptogenesis; 
neutrino oscillations; 
introducing an independent source of $CP$ violation in $N_1$ decays;
partial loop cancellations; 
supersymmetry, and;
reducing the (possibly effective) vev entering the seesaw.
We showed how to construct viable, natural $\nu$2HDMs
which utilise the latter mechanism \cite{Clarke2015hta}.
Such models predict an automatically SM-like Higgs boson, 
(maximally) TeV-scale scalar states, 
and low- to intermediate-scale hierarchical leptogenesis 
with $10^3\text{ GeV}\lesssim M_{N_1}\lesssim 10^7\text{ GeV}$.
One version (the radiative Ma model)
also includes a dark matter candidate.

\acknowledgments

This work was supported in part by the Australian Research Council.
Figures are reproduced from \cite{Clarke2015gwa,Clarke2015hta}.
I would like to thank the conference organisers for their hospitality.

\bibliographystyle{JHEP.bst}
\bibliography{references}

\end{document}